# Inversion-free, noiseless Raman echoes

Byoung S. Ham
*Center for Photon Information Processing, School of Electrical Engineering, Inha University*
*253 Yonghyun-dong, Nam-gu, Inchoen 402-751, S. Korea*
bham@inha.ac.kr

**Abstract:** Using double optical Raman rephasing, an inversion-free resonant Raman echo is studied in an inhomogeneously broadened spin ensemble of a solid medium, where the Raman optical field-excited spin coherence has a frozen propagation vector. Unlike photon echoes whose quantum memory application is strictly limited due to π rephasing pulse-induced population inversion causing quantum noises, the optical Raman field-excited spin echo is inherently silent owing to the frozen propagation vector. Thus, the doubly rephased Raman echo can be directly applied for quantum interface in a population inversion-free environment.
PACS: 42.50.Md, 82.53.Kp

Quantum interface between photons and a matter is an essential step for practical quantum information processing using flying qubits of photons [1]. In this context, quantum memory has been intensively studied recently in solid media [2-11]. Compared with single atom-based quantum memories [2,3], ensemble-based quantum memories offer benefits of multimode and wide bandwidth [4-11]. Strong absorption in an ensemble is another important factor for the quantum interface to provide a maximum fidelity at a dense information rate. Thus, photon echoes in an optically thick ensemble carry intrinsic benefits compared with a single atom-based technique. However, photon echoes have a critical drawback of π rephasing pulse-induced population inversion, where the population inversion is a potential source of spontaneous and stimulated quantum noises [12]. Moreover, photon echo efficiency in most solid media is normally extremely low to be around 1% due to echo reabosrption in a forward propagation scheme [13]. A backward photon echo scheme using optical locking has been demonstrated recently to eliminate the echo reabsorption [14]. To solve the population inversion problem, a double rephasing technique has recently been suggested for silent echo generation, where the silent echo becomes a source of the inversion-free echo [15,16].

Recent demonstration of Hahn echo (or spin echo) in a single nitrogen-vacancy (NV) center of diamond [3] shows extremely high retrieval efficiency due to an extremely strong oscillator strength compared with most rare-earth doped solids [17]. Although spin decoherence is robust compared with the optical counterpart, Hahn (spin) echoes suffer from lower resolution (λ~10 cm) and limited bandwidth (δλ~$10^4$ Hz) [3]. In (resonant) Raman echoes, an optical Raman pulse is used for rephasing of the spin ensemble, where resolution and bandwidth is several orders of magnitude higher as in photon echoes [18-21]. Compared with Hahn echoes utilizing a two-level spin transition, the Raman optical echo uses a three-level optical medium to excite the spin coherence indirectly. Thus, Raman echoes offer benefits of higher optical resolution as in photon echoes and slower decoherence rates as in Hahn echoes [21]. Moreover, controlling Raman optical fields for a backward propagation scheme is an essential benefit for enhanced echo efficiency. Even in a weak field limit, efficient Raman coherence excitation has been demonstrated decades ago [22], and very high Raman coherence retrieval efficiency has also been demonstrated [8,21].

Decades ago, a prototype of the resonant Raman echoes was experimentally demonstrated [20,21] and analyzed recently [23], where the data pulse duration is much shorter than the spin inhomogeneous broadening, but the storage time is much longer than the optical decay times. An optical locking technique has also been applied to the Raman echoes to control optically excited atomic population decay process [24]. However, the population inversion constraint in previous Raman echo models has not been resolved yet. Here, we present a population inversion-free Raman echo scheme by using double rephasing for quantum memory applications. Just like conventional photon echoes [12], the phase matching condition in the Raman echo scheme is sustained only for the optical transitions, where spin coherence excitation has no effect due to frozen propagation vectors [25]. Thus, the Raman optical field-excited spin echo becomes inherently a silent echo, where it does not alter system coherence. This silent echo leads to the inversion-free, noiseless Raman echo by another Raman rephasing pulse.

Figure 1 shows a schematic of a doubly rephased Raman echo model. The Raman pulses $\Omega_P$ and $\Omega_C$ in Fig. 1(a) compose each Raman pulse D, R1, and R2 in Fig. 1(b). The phase matching condition of the propagation vectors applies only for the optical components, where $\Omega_{P'}$ represents the Raman echo-converted optical signal by the optical readout pulse $\Omega_{C'}$: E2 (E1) is without (with) population inversion (see Fig. 4). Thus, the present model also implies a near complete image processing as its importance is indicated in quantum imaging [26].

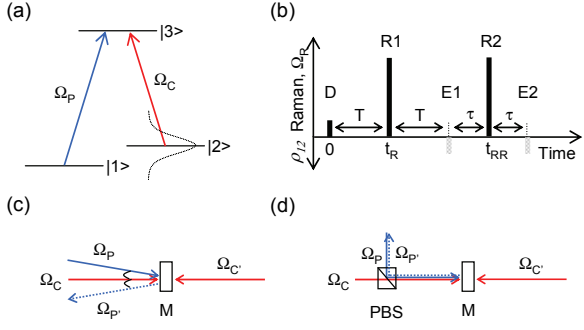

**Fig. 1.** Doubly rephrased Raman echo. (a) Energy level diagram for Raman fields. (b) Pulse sequence. Each pulse (D, R1, and R2) is composed of the Raman pulse of $\Omega_P$ and $\Omega_C$ in (a). E1 and E2 are the Raman echoes with and without population inversion, respectively. (c) Noncollinear and (d) Collinear propagation scheme. $\Omega_{P'}$ is the signal read by $\Omega_{C'}$, where $\Omega_{C'}$ is a readout pulse for Raman echo E2. M: optical medium. PBS: polarized beam splitter.

In Fig. 2, we discuss coherent population transfer in a lambda-type three-level optical system composed of two ground states (|1> and |2>) and an excited state (|3>) as shown in Fig. 1(a). For the numerical calculations we solve time-dependent density matrix equations assuming that all atoms are Gaussian distributed in both optical and spin transitions. The main density matrix equation for the Raman excitation is:

$$\frac{d\rho_{12}}{dt} = -i\Omega_C \rho_{13} + i\Omega_P \rho_{32} - (i\delta + \gamma)\rho_{12}, \quad (1)$$

where $\Omega_P$ and $\Omega_C$ compose a Raman pulse, $\delta$ is a two-photon detuning of the Raman pulse, and $\gamma_{12}$ is a spin phase decay rate. However, only two-photon detuning effect is applied for the calculations because optical broadening simply contributes to the line narrowing of the Raman coherence excitation [27]. Figures 2(b)~2(d) show a weak field limit case, where the Raman optical pulse is composed of a probe ($\Omega_P$=50 kHz) and a coupling ($\Omega_C$): $\Omega_P$ is resonant to the optical transition of |1>–|3>; and $\Omega_C$ is resonant to the other optical transition of |2>–|3>.

Figure 2(b) shows for a generalized Rabi frequency $\Omega_R$=200 kHz, where $\Omega_R^2 = \Omega_C^2 + \Omega_P^2$, and the excited Raman coherence is damping out quickly due to detuned atoms (or spins) by $\delta$ from the resonance frequency: See the inhomogeneously broadened spin ensemble in Fig. 2(a). Figure 2(c) shows coherence evolutions of a perfectly resonant atom group with $\delta$=0 in Fig. 2(b), where a complete population transfer occurs at $\Omega_R \cdot \Delta T = 2(2n-1)\pi$, and $\Delta T$ (5 µs; dash-dot line) is the pulse duration of $\Omega_R$ (0.2 MHz multiplied by $2\pi$). The Raman excited spin coherence (Re$\rho_{12}$) is maximized when a perfect population transfer occurs. At the same time, optical coherence (Im$\rho_{13}$) and excited state population ($\rho_{33}$) become zero as expected from Eq. (1). This means that the optical readout process for the Raman coherence must be noise-free, as we discuss in Fig. 3. For a detuned atom group as shown in Fig. 2(d), however, all effects are damped out due to modified $\Omega_R$ by $\delta$.

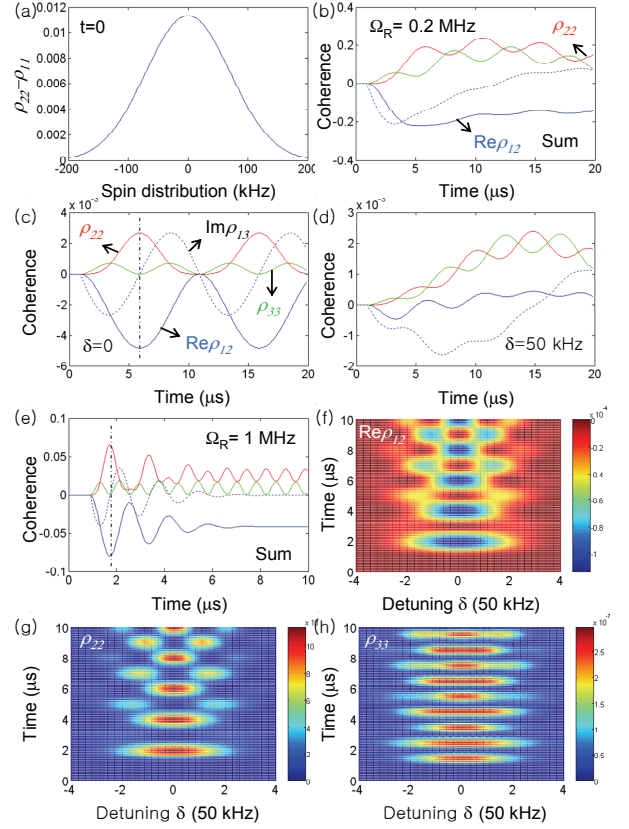

**Fig. 2.** Raman field excited spin coherence. (a) Spin inhomogeneous broadening between states |1> and |2>. (b)–(d) $\Omega_P$=50 kHz, $\Omega_R$=200 kHz; $\Omega_R^2=\Omega_P^2+\Omega_R^2$. (e)-(h) $\Omega_P$=50 kHz, $\Omega_R$=1 MHz. Red: $\rho_{22}$; Blue Re$\rho_{12}$; Dotted: Im$\rho_{13}$; Green: $\rho_{33}$. Detuning $\delta$ is for two-photon detuning for $\Omega_P$ and $\Omega_C$. All decay rates are zero. Initial atom population is $\rho_{11}$=1.

With a coupling Rabi much greater than the atom broadening in Fig. 2(e), the overall atom ensemble behaves like the resonant atom group as shown in Fig. 1(c). Figs. 2(f)~2(h) represent Raman coherence and population behaviors of all atom groups, where the atom detuning effect appears after a few cycles of Rabi flopping. Thus, maximum Raman coherence excitation can be obtained easily by controlling the coupling Rabi frequency $\Omega_C$.

Figure 3 represents the present noise-free Raman echo based on double Raman rephasing. In the inset of Fig. 3(a), D, R1, and R2 are Raman pulses composed of $\Omega_P$ and $\Omega_C$. For the Raman rephasing pulses R1 and R2, $\Omega_P$ is the same as $\Omega_C$ to satisfy a complete population swapping between states |1> and |2> [23]. The first Raman echo E1 is under population inversion as shown in Fig. 3(b). However, the second echo E2 by the double Raman rephasing results in no population inversion, where final $\rho_{22}$ is the initial $\rho_{22}$ excited by D. As addressed in the introduction, the Raman optical field-excited coherence $\rho_{12}$ is frozen for propagation. Thus, unlike photon (or Hahn) echoes satisfying phase matching, the echo E1 does not contribute to stimulated emission even under massive population inversion ($\rho_{22} \gg \rho_{11}$). This silent echo E1 becomes the origin of the inversion-free echo E2. With no optical population ($\rho_{33}=0$), E2 becomes a noiseless Raman echo.

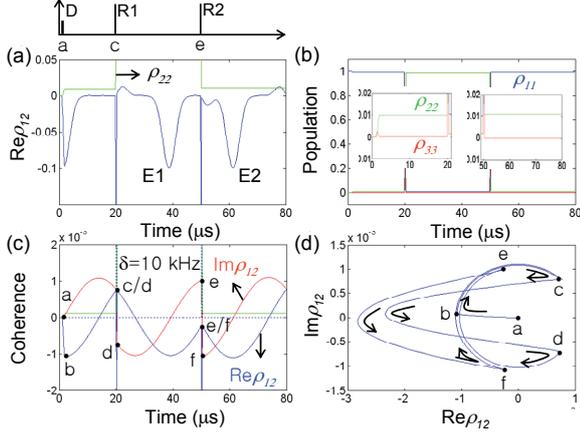

**Fig. 3.** Inversion-free Raman echo using double Raman rephasing. (a) E1 (E2) represents a Raman echo with (without) population inversion. Each R1 and R2 satisfies a $2\pi$ pulse area, where $\Omega_P=\Omega_C=2.5/\sqrt{2}$ MHz multiplied by $2\pi$ with a pulse duration of 400 ns. For D, $\Omega_R=1$MHz ($\Omega_P=50$ kHz) with a Raman pulse duration of 1 $\mu$s. (b) Population evolution for (a). Blue: $\rho_{11}$; Green: $\rho_{22}$; Red: $\rho_{33}$. (c) Coherence evolution for a detuned atom group in (a). a-b: D; c-d: R1; e-f: R2. (d) Bloch vector model for (c). All decay rates are zero. Initial atom population is $\rho_{11}=1$.

Figure 3(c) shows the phase shift of $\rho_{12}$ when Raman rephasing is applied at t=20 $\mu$s and at t=50 $\mu$s. Like Hahn echoes, the coherence $\rho_{12}$ shows a mirror image-like behavior with a $\pi$ phase shift across the Raman rephasing pulse, representing a time-reversal process. Figure 3(d) represents a Bloch vector model for Fig. 3(c). Compared with Hahn echoes, where the uv plane flip-over occurs along the u-axis (Re$\rho_{12}$), the Raman rephasing occurs along the v (Im$\rho_{12}$) axis: For a detailed analysis, see ref. 23. This difference is the direct result of two-photon coherence with a $\pi$ phase shift [23].

Now we discuss a readout procedure, where the Raman coherence retrieval (Raman echo) must be converted into optical coherence for photon generation to conclude the quantum interface. Because phase matching is satisfied only for the optical components in Raman transitions, a phase conjugate scheme can be obtained if all beams are collinear, and C1 and C2 counterpropagate (see Fig. 1(d)):

$$k_E = k_P - k_{C1} + k_{C2}, \quad (2)$$
$$\omega_E = \omega_P - \omega_{C1} + \omega_{C2},$$

where $k_X$ and $\omega_X$ are propagation vector and frequency of the pulse $X$, respectively: $\mathbf{k_P}$ is for $\Omega_P$ in D. E is the C2 converted optical coherence (Im$\rho_{13}$) for the Raman echo E2 at time $T_E$ (see the red curve at t~60 $\mu$s in Fig. 4(a)):

$$T_E = 2T_{R2} - T_{E1}, \quad (3)$$

where $T_X$ is the time of the pulse $X$. For maximum readout signal E, the pulse area of C2 needs to be adjusted for a complete depletion of the Raman echo E2 as shown in Fig. 4(b). In the implementation, however, every transferred $\rho_{33}$ must turn into a phase conjugate $\Omega_E$, satisfying Eq. (2), where there is only Raman coherence depletion regardless of C2 pulse length [19]. This phase conjugate scheme is important for a nearly perfect quantum interface, where the optical signal $\Omega_E$ retraces backward along the data path ($\mathbf{k_P}$(D)) without absorption [28].

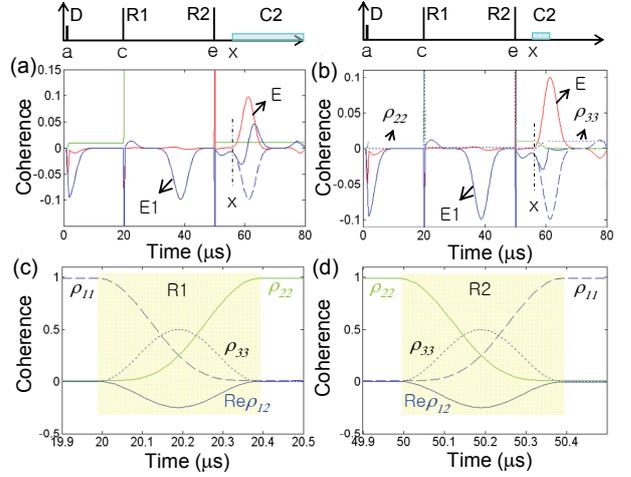

**Fig. 4.** (a) and (b) Optical readout E (red) of the Raman echo E2 (dashed) by C2 at $\Omega_{C2}=100$ kHz. x is the switch-on time for C2. Dashed: E2 without C2. (c) and (d) Population swapping by R1 and R2, respectively. All parameters are the same as in Fig. 3 unless otherwise noted. The yellow shaded region represents the pulse area of Raman rephasing R1 or R2.

Figures 4(c) and 4(d) represent the function of the optical Raman pulses R1 and R2. As shown, the $2\pi$ Raman pulse results in a complete population swapping between $\rho_{11}$ in state |1> and $\rho_{22}$ in state |2> via the excited state |3>. See that no optical population $\rho_{33}$ is left in state |3> by R1 or R2. Thus, the optical coherence E in Figs. 4(a) and 4(b) results only from the excited coherence $\rho_{12}$ by $\Omega_P$ in the data Raman pulse D via Raman rephasing, which means that the phase conjugate signal becomes noise free.

In conclusion, a double optical Raman rephasing scheme was presented for an inversion-free, noiseless Raman echo owing to a frozen propagation vector of the Raman optical field-excited spin coherence. By controlling one leg of a data Raman pulse, nearly perfect atom population transfer between two ground states can be achieved for a maximum Raman coherence excitation. By controlling a readout optical pulse propagation direction satisfying a phase conjugate scheme, nearly perfect quantum interface can be also obtained with noise-free Raman conversion.

**Acknowledgment**
This work was supported by the Creative Research Initiative Program (grant no. 2011-0000433) of the Korean Ministry of Education, Science and Technology via the National Research Foundation.